\documentclass[pdflatex]{sn-jnl}

\usepackage{appendix}
\usepackage{siunitx}
\usepackage{algorithmicx}
\usepackage{algorithm}
\usepackage{algpseudocode}
\usepackage{amsmath}
\usepackage{amssymb}
\usepackage{multirow}
\usepackage[dvipsnames]{xcolor}
\usepackage{comment}
\usepackage{etoolbox}
\usepackage{xspace}
\usepackage{booktabs}
\usepackage{hhline}

\makeatletter
\providecommand{\cline}[1]{\@cline{#1}}
\makeatother

\def\tsc#1{\csdef{#1}{\textsc{\lowercase{#1}}\xspace}}
\tsc{WGM}
\tsc{QE}

\begin{document}

\title{Predicting Company Growth using Scaling Theory Informed Machine Learning}

\author[1,2]{\fnm{Ruyi} \sur{Tao}}\email{taoruyi@mail.bnu.edu.cn}
\author[3]{\fnm{Veronica R.} \sur{Cappelli}}
\author[1,2]{\fnm{Kaiwei} \sur{Liu}}
\author[4,5]{\fnm{Marcus J.} \sur{Hamilton}}
\author[4]{\fnm{Christopher P.} \sur{Kempes}}
\author[4]{\fnm{Geoffrey B.} \sur{West}}
\author*[1,2]{\fnm{Jiang} \sur{Zhang}}\email{zhangjiang@bnu.edu.cn}

\affil[1]{\orgdiv{Department of Systems Science}, \orgname{Beijing Normal University}, \orgaddress{\city{Beijing}, \country{China}}}
\affil[2]{\orgname{Swarma Research}, \orgaddress{\city{Beijing}, \country{China}}}
\affil[3]{\orgdiv{Department of Managerial Decision Sciences}, \orgname{IESE Business School}, \orgaddress{\city{Barcelona}, \country{Spain}}}
\affil[4]{\orgname{The Santa Fe Institute}, \orgaddress{\city{Santa Fe}, \country{USA}}}
\affil[5]{\orgdiv{Department of Anthropology and School of Data Science}, \orgname{University of Texas at San Antonio}, \orgaddress{\city{San Antonio}, \country{USA}}}

\abstract{Predicting company growth is a critical yet challenging task because observed dynamics blend an underlying structural growth trend with volatile fluctuations. Here, we propose a Scaling-Theory-Informed Machine Learning (STIML) framework that integrates a scaling-based growth model to capture the mechanism-driven average trend,  together with a data-driven forecasting model to learn the residual fluctuations. Using Compustat annual financial statement data (1950--2019) for 31,553 North American companies, we extend the growth model beyond assets to multiple financial indicators, and evaluate STIML against growth model-only and purely data-driven baselines. Across 16 target variables, we show that company growth exhibits a clear separation between trend-driven predictability and fluctuation-driven predictability, with their relative importance depending strongly on company size and volatility. Interpretability analyses further show that STIML captures multivariate dependencies beyond simple autocorrelation, and that macroeconomic variables contribute significantly less to predictive performance on average. Moreover, we find the scaling-based growth model overlooks asymmetric deviations, which instead contain the structured and learnable signals, suggesting a path to refine mechanistic models.}


\maketitle

\section{Introduction}
\label{sec:Introduction}

Companies are fundamental units of modern economic activity. Accurate and interpretable prediction of company growth facilitates the assessment of future performance, thereby providing valuable guidance for strategic decision-making, risk management, and related fields\cite{Sterk2021}. Moreover, companies are complex systems, characterized by navigating diverse internal information flows and interactions with dynamic externalities, and displaying characteristics of complex systems such as nonlinearity, path dependence, emergence, modularity, and scale-dependence\cite{arthur1994increasing}. Understanding and modeling company growth is not only of practical economic significance but also advances theoretical understanding of growth in complex systems. However, due to the inherent complexity of company dynamics--such as rapid changes in the environment, the high sensitivity of companies to external factors, the heterogeneity of companies, and the limited understanding of its growth mechanisms--achieving both accuracy and interpretability in company growth prediction remains a challenge.

Company growth encompasses expansion in size (e.g., assets, employees, revenue) as well as increases in net profit, market share, stock price, and market value\cite{DELMAR2003189, Brush2000, Dang2018}. Any quantitative data-driven perspective on company growth must be built on the foundation of financial indicators.
Derived from accounting records, financial indicators offer a structured, empirical view of company operations and form a basis for assessing growth. Building on this role, they support a wide range of downstream tasks, including asset pricing and valuation, bankruptcy prediction, risk assessment, and fundamentals-based stock forecasting\cite{su14020770,ALAKA2018164,georgieva2009analysis}. 
This study aims to generate comprehensive forecasts of financial indicators to support accurate growth predictions.

Forecasting financial indicators is a time series problem that aims to predict future values based on historical observations. In recent years, time series forecasting has become increasingly important, with broad and effective application across many fields\cite{Masini2023}. 
Within financial indicator forecasting, existing studies have focused primarily on revenues, sales, and net profit\cite{Barker2018,Xinyue2020,ZOGRAFOPOULOS2025257,Ding2020,Lee2017}. Yet, most of this work emphasizes prediction accuracy, with limited attention to the underlying mechanisms governing company growth. A central question remains: is company growth largely an unpredictable stochastic process dominated by random fluctuations, or does it follow systematic patterns that can be identified and predicted? Without addressing these fundamentals, time series forecasting methods—especially high-capacity models such as neural networks—may mistake noise for signal, leading to overfitting and unstable predictions\cite{Yang2020,Casolaro2023}.


The debate over whether company growth follows systematic patterns can be traced back to Gibrat’s law in the 1930s\cite{Sutton1997}. This law posits that company growth is a purely random process, where growth rates are independent of firm size. This hypothesis sparked extensive empirical research, yet no universal conclusion has been achieved\cite{Daunfeldt2011,Coad2010,coad2007firm}. 
Following empirical studies have shown that firm sizes follow Zipf’s law, exhibiting a stable power-law distribution across different economies and time periods\cite{axtell2001zipf,GABAIX2011}. 
At the same time, Stanley and other econophysicists demonstrated that growth rate fluctuations exhibit systematic scaling with firm size, revealing universal statistical regularities in company growth dynamics\cite{Stanley1996,STANLEY1999156,Jakovac2020}.
Together, these findings suggest that although company growth exhibits stochastic fluctuations, it also follows underlying inherent patterns \cite{Fu2005}.  Motivated by these observations, subsequent studies proposed a range of growth models, such as the granular model\cite{GABAIX2011,Moran2024}, which treats companies as aggregates of microscopic particles. However, these studies primarily focused on reproducing macro-level statistical distributions, emphasizing fluctuations while largely neglecting the evolution of the average growth trend.

To address this gap, Zhang and colleagues \cite{zhang2021scaling} established a mechanistic differential equation for company growth,  on the basis of the scaling laws and financial balance equation. By conducting empirical analyses of the U.S. and Chinese markets, they found that this model successfully captured the average growth of companies.  This work clarifies the mechanisms of company growth and establishes a predictive framework grounded in scaling theory.
Nevertheless, 
from the perspective of predicting growth at the individual-company level, there is still considerable room for improvement in these models. For example, after modeling the average trend, the relative prediction errors for assets remain around 20\%. The original study reported that 
the residuals of this model may remain correlated with firm size, if so, suggesting that unexplained fluctuations may still contain predictable structure.
These may arise partly from modeling approximations that omit variables related to company growth,
and partly from the inherent complexity of individual companies driven by interactions among multiple financial factors. 
In addition, the original growth equation focuses mainly on company assets, while a full assessment of company status requires forecasting a broader range of financial indicators.

To predict financial indicators at the company level, we propose a framework that explicitly models \textit{both} the average trend and fluctuations of company growth. 
This yields a hybrid prediction framework that integrates mechanism-based and data-driven modeling. We refer to this approach as Scaling-Theory-Informed Machine Learning (STIML). {Our model outperforms benchmark approaches and, beyond prediction, our analysis \iffalse connects quantitative growth modeling with classical management theories and \fi provides system-level insights for company modeling}. Our contributions can be summarized as follows:
\begin{enumerate}

    
    

    \item  We propose a Scaling-theory-informed Machine Learning (STIML) framework that integrates {extended growth model (GM), building on the growth model proposed by \cite{zhang2021scaling},} to capture average trend, with machine learning (ML) of residual fluctuations, enabling improved predictive accuracy while preserving interpretability.

    \item We identify two distinct components of company growth—trend predictability and fluctuation predictability—whose relative importance varies with firm size and volatility. The GM performs best for large and stable firms, where growth is dominated by structural trends,  while ML better captures predictable fluctuations in small firms. By explicitly modeling both trend and fluctuations, STIML improves accuracy across firm sizes. This reveals a regime-dependent structure of predictability in company growth.

    \item {After conducting feature importance analysis,  we find that the STIML also uncovers interactions among financial variables beyond autocorrelation. And we also find that macroeconomic variables contribute little to overall prediction, a pattern robust across company sizes. Moreover, compared with small companies, the fluctuations of large companies’ financial indicators rely more heavily on autocorrelation, reflecting greater organizational inertia.}

    \item {As a hybrid model, the performance gains of STIML highlight the limits of both pure ML and GM. Compared with pure ML, STIML extracts predictable structure missed by ML alone,  leading to consistent accuracy improvements across firm sizes. Relative to GM, STIML shows that deviations from scaling laws are not just noise but contain learnable predictive signals, particularly under negative shocks where systematic asymmetry emerges. This enables systematic error correction in the negative-residual regime and suggesting a path to refine mechanistic models.}


\end{enumerate}

\section{Methodology}\label{sec:methods}

The central premise of our work is to divide the growth of a company into two components, namely, \emph{mechanism-based growth} and \emph{fluctuation}. In the following sections, we provide a more comprehensive description of modeling these two components, unraveling their technical details and interplay in the broader context of company growth.

\begin{figure}
    \centering
    \includegraphics[width=1\linewidth]{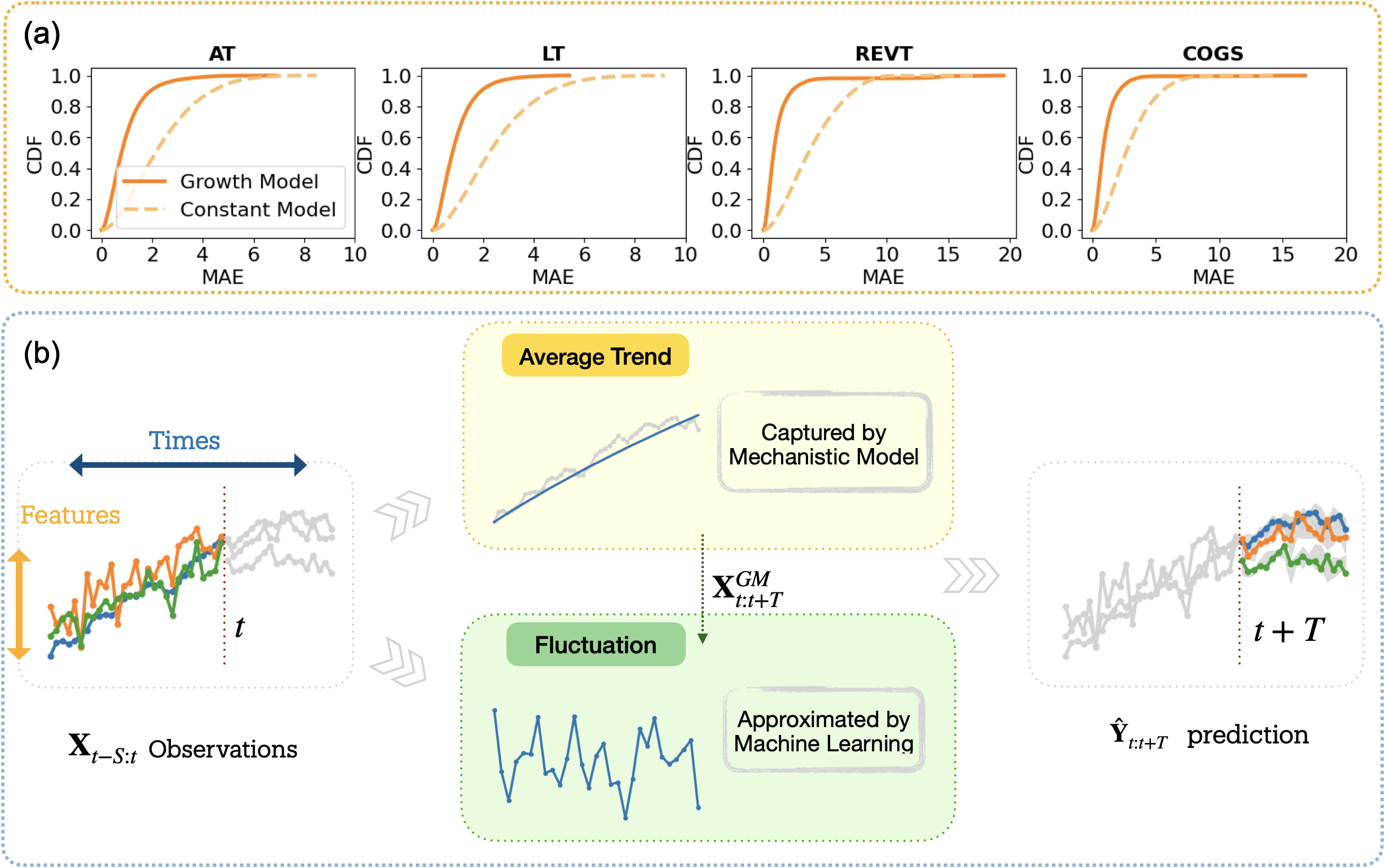}
    \caption{\textbf{Performance of the extended GM and the STIML framework.}(a). The cumulative distribution of MAE for the extended GM and the constant model. (b). The framework of STIML. The model consists of two components: the mechanistic model captures the average trend of company growth, while time-series prediction techniques capture the fluctuations. By combining both components, the final prediction of future growth is obtained.\iffalse shows prediction accuracy of GM parts compared to the baseline-constant model. (b) is the conceptual framework of our model. Combining the growth model and time series prediction techniques, we can obtain the prediction trajectory in the future time step $t$ to $t+T$ according to the company trajectory observation. \fi}
    \label{fig:framework}
\end{figure}

\subsection{Problem Definition} 
{For multivariate time series forecasting tasks, given historical observations $\mathbf{X}_{t-S:t} = \{ \mathbf{X}_{t-S}, \dots, \mathbf{X}_t\} \in \mathbb{R}^{S \times N'}$ with $S$ time steps and $N'$ variates, we predict the future $T$ time steps and $N$ variates, that is $\mathbf{Y}_{t+1:t+T} = \{\mathbf{X}_{t+1}, \dots, \mathbf{X}_{t+T}\} \in \mathbb{R}^{T \times N}$. In general, $N'$ input variables must contain $N$ predicted variables. For convenience, we denote $\mathbf{X}_{t,:}$ as the simultaneously recorded time points at time step $t$, and $\mathbf{X_{:,n}}$ as the whole time series of each variate indexed by $n$ (same as $\mathbf{Y_{t,:}}$ and $\mathbf{Y_{:,n}}$). }

\subsection{Company Growth Model}\label{sec:growth_model}

Zhang et al. \cite{zhang2021scaling} propose a company growth model by deriving a mechanistic differential equation for firms’ asset growth from the scaling law, which we refer to as the Growth Model (GM). The GM is as follows:
 
\begin{equation}
    \label{eq:growth_eq}
    \frac{dA}{dt} = \frac{c_IA^{\beta_I}}{1-c_L\beta_LA^{\beta_L-1}}
\end{equation}

In the equation above, $A$ is the total assets of a representative firm, $\beta_L$ and $c_L$ are the scaling parameters of the exponent and constant for the power law relationship between liability($L$) and assets, i.e. $L = c_L A^{\beta_L}$. And $\beta_I$ and $c_I$ are for the scaling between net profit($I$) and assets ($I = c_I A^{\beta_I}$). Equation \ref{eq:growth_eq} outlines the growth of a company from first-principles perspective, lending an intuitive and comprehensive understanding of the company's growth dynamics. 
The model offers an approach to understanding the growth trajectory of an idealized company, tying together the relationships between various financial indicators and the influence of net profit and liability on a company's asset growth. This equation omits investment, dividends, and other growth terms because their magnitudes are much smaller than net profit, total assets, and liabilities in the empirical data. Additionally, experiments in \cite{zhang2021scaling} show that excluding them does not affect the predictability of Equation \ref{eq:growth_eq}. 

 %


Here, to enable predictions for a broader set of financial indicators rather than only assets as in the original GM,  we extend Equation \ref{eq:growth_eq} to other financial variables $x$, such as liabilities, revenues, cost. This extension is based on the empirical scaling relationship between each financial indicator $x$ and assets $A$, expressed as $x=c_xA^{\beta_{x}}$ (see Figure \ref{fig:scaling}). Because $x$ denotes multiple financial indicators, we substitute their asset-scaling laws into the original growth equation (Equation \ref{eq:growth_eq}) to obtain generalized scaling-based growth equations—one for each variable of interest:

\begin{equation}
    \label{eq:growth_eq_extension}
    \frac{dx}{dt}
    = \frac{dx}{dA} \frac{dA}{dt} = \frac{c_x c_I \beta_x A^{\beta_x + \beta_I -1}}{1-c_L\beta_LA^{\beta_L-1}}
\end{equation}

The solution to the above equation is the GM predictor, {which we denote $x^{GM}$}. $c_x$ and $\beta_x$ are scaling parameters estimated from training data.  
Therefore, for every financial indicator $x$ that has a power law relationship with assets $A$, we can calculate its mechanism-based growth. 


{To implement this procedure, we begin with the scaling relation and take logs of $x$ and $A$, yielding $\ln x = \ln c_x + \beta_x \ln A$. We then fit a linear model to the training set using ordinary least squares (OLS) to estimate $c_x$ and $\beta_x$ for each attribute. We subsequently compute the predicted result {$\mathbf{X}^{GM}_{t:t+T,:}= \{\mathbf{X}^{GM}_{t+1}, \dots, \mathbf{X}^{GM}_{t+T} \} \in \mathbb{R}^{T \times N}$} via Equation \ref{eq:growth_eq_extension} with the Euler method {for a given initial value $\mathbf{X}_t$}, where $\mathbf{X}_t$ denotes the $N$ dimensional vector of variable at time $t$. All the data used in our analysis are annual financial indicators, and in our model, time $t$ represents one financial reporting year.

{We calculate the mean absolute error (MAE) between mechanism-based growth from Equation \ref{eq:growth_eq_extension} and the actual value for all the test companies, and }Figure \ref{fig:framework} (a) shows the cumulative MAE distribution  compared with a constant model  
based on Gibrat's Law\cite{Sutton1997}, which assumes that growth rates are independent of size. This is a standard baseline for company growth. We extend the constant model to other indicators as additional baselines. Across all indicators, Figure \ref{fig:framework} (a) shows GM’s superior performance.

\subsection{Constructing Hybrid Forecasting Models}\label{sec:TS_model}

Although the GM is designed to capture average trends, it cannot adequately account for the many factors driving the observed heterogeneity in company growth. For this reason, we introduce a machine learning model framework called Scaling Theory Informed Machine Learning (STIML) to predict observed fluctuations in individual company growth. 
{In particular, our approach draws from the physics informed machine learning (PIML) method}.
The key of PIML lies in integrating known physical knowledge as a bias to design the model's architecture or training process. 
Therefore, STIML revised the machine learning model in two ways: 

\begin{enumerate}
    \item The predictions generated by the growth model are also included as part of the model's input to guide future predictions. Specifically, our model takes two kinds of inputs: one is the historical sequence of the target variable $\mathbf{X}_{t-S:t,:}$, and the other is the future $T$ steps predictions $\mathbf{X}^{GM}_{t:t+T,:}$  from the GM (Equation \ref{eq:growth_eq_extension}). This can serve as a heuristic signal for the future.

    \item In STIML, the target that the integrated machine learning model to predict is the difference between the true values and the predictions of GM, that is $\mathbf{Y} - \mathbf{X}^{GM}$ , rather than directly predicting the true value $\mathbf{Y}$, to reduce the difficulty of prediction. 
\end{enumerate}




In this study, we integrate several representative time-series forecasting models into the STIML framework, including classical methods such as Random Forest (RF) and VARIMA, as well as neural network–based approaches. For the latter, we selected a traditional multilayer perceptron(MLP) and the iTransformer\cite{liu2024itransformer}, a state-of-the-art model that has shown good performance in multivariate forecasting tasks. We denote by GM-X a hybrid STIML model that integrates the scaling based GM with the machine learning model X.

\textbf{GM-RF and GM-VARIMA}

For both the RF and VARIMA models, we adopt readily available implementation libraries. The input to these models is the sequence ${\mathbf{X}_{t-S:t,:}}$, which includes the target variables ${\mathbf{X}_{t-S:t,:N} \in \mathbb{R}^{S \times N}}$ as well as additional potentially informative predictors,  such as macroeconomic information $\mathbf{X}^{macro}_{t-S:t,:} \in \mathbb{R}^{S \times (N'-N)}$. These features are concatenated to form the final input.  The prediction target is defined as the residual between the observed values and the growth model prediction, i.e.,$\mathbf{Y} - \mathbf{X}^{GM}$.


\textbf{GM-MLP and GM-iTransformer}

For the neural network models, we provide a more detailed description of the architecture and implementation to better tailor it to the specific requirements of our task.

Our neural network model adopts a classic encoder-decoder architecture, which offers the flexibility to adjust the number of input variables $N'$, the input sequence length $S$, as well as the number of output variables $N$ and output sequence length $T$. In this setup, the model inputs include a company's financial indicators along with other relevant features that may support prediction, such as macroeconomic variables. The outputs correspond to the target financial indicators of the company. The detailed data preprocessing could be found in Section \ref{sec:datapreprocess} and \ref{sec:pre_progress_si}.

Here, we selected two typical neural network models, MLP and iTransformer, as the representatives of neural network, and named them GM-MLP, GM-iTransformer, respectively. 
MLP is a classic and simple neural network model and, in theory, serves as a universal approximator capable of fitting any complex function. Meanwhile, iTransformer, as a state-of-the-art model for multivariate time series forecasting tasks, leverages attention mechanisms to capture long-range dependencies and construct attention matrices between variables to model interdependencies among multiple variables\cite{liu2024itransformer}.

Model details can be found in Section \ref{sec:itrans_arch}, and we can obtain the final prediction $\mathbf{\hat{Y}}_{t:t+T} = \mathbf{O}_{t:t+T} + \mathbf{X}^{GM}_{t:t+T}$, where $\mathbf{O}_{t:t+T}$ is the output of NN.  The loss function here is the mean square error (MSE), as it generally provides superior performance in most cases. It optimizes the difference between the prediction $ \mathbf{\hat{Y}} $ and the observed data $\mathbf{Y}$ in log space.

\section{Data}\label{sec:datapreprocess}

We use the Compustat datasets, encompassing nearly 70 years of annual financial statement data from 31,553 publicly traded North American companies, spanning from 1950 to 2019. The features extracted are primarily from the three core financial statements. Additionally, we gathered macroeconomic data for the U.S. since 1950 to 2019 from Compustat North America and Compustat Historical databases compiled by Standard \& Poor’s \cite{SP}.

{In the data preprocessing stage, we applied feature selection, noise reduction, missing value imputation, inflation adjustment, and logarithmic transformation to enhance the learnability of the data. Detailed procedures are provided in the Appendix (Section \ref{sec:pre_progress_si}).}

We retained 23 financial features and 12 macroeconomic features for 26,038 companies, which constitute our entire dataset. The dimensionality of the input for the encoder module $N' = 35$ at most. We selected these indicators as they provide a comprehensive view of the financial performance and economic environment associated with each company. The detailed features are shown in Table \ref{tab:finance} and Figure \ref{fig:macro_index}. These features were selected to encompass a wide range of financial health indicators, such as size,  profitability, liquidity, and leverage, along with key economic factors that may influence a company's performance, such as GDP, inflation consumer prices, etc. Our aim was to build a dataset robust enough to capture the multifaceted nature of company growth and its influencing factors. 

{For our model, we focus on predicting financial attributes that exhibit scaling relationships with company assets. Figure \ref{fig:scaling} illustrates the empirical relationship between these attributes and assets. Among the 23 financial features, we find that 15 of them show clear scaling behavior with respect to total assets—excluding the 7 features that may take negative values (see Table \ref{tab:finance}). This implies that, including assets itself, our dataset allows for the simultaneous prediction of up to 16 financial indicators that demonstrate reliable scaling behavior.}

{Details of the data setup and hyperparameter choices are provided in Section~\ref{sec:pre_progress_si}.}

\section{Results}\label{sec:experiments}

\subsection{Prediction Accuracy}\label{sec:acc}

To validate the effectiveness of our framework, we compared our model against several benchmark models, including an average baseline model, the Growth Model (GM), and a purely data-driven machine learning model (ML). Table \ref{tab:compare_models} summarizes the key differences among these baselines. This comparison not only evaluates predictive performance but also helps to clarify how our framework works.  {In all experiments, we use mean absolute error (MAE) in log space as the evaluation criterion. This metric approximates relative error when prediction errors are small; that is, an MAE of 0.1 is approximately equivalent to a relative error of 10\%.}

\begin{table}[]
    \centering
    \begin{tabular}{c|p{10cm}}
        \hline
         \textbf{Model}  & Description  \\
         \hline
          Baseline & Simple moving average model $\frac{1}{S}\sum_i\mathbf{X}_i$, where $S$ is the history time length.\\
          \hline
          GM &  Applying the Euler's method to compute the predicted value $\mathbf{X}_{t:t+T}$  base on the Equation \ref{eq:growth_eq_extension}, with the initial value $\mathbf{X}_t$. \\
          \hline
          ML &  The optimization objective is directly $\mathbf{Y}$ \\
        \hline
    \end{tabular}
    \caption{The Comparative Models}
    \label{tab:compare_models}
\end{table}


\begin{figure}
    \centering
    \includegraphics[width=1\linewidth]{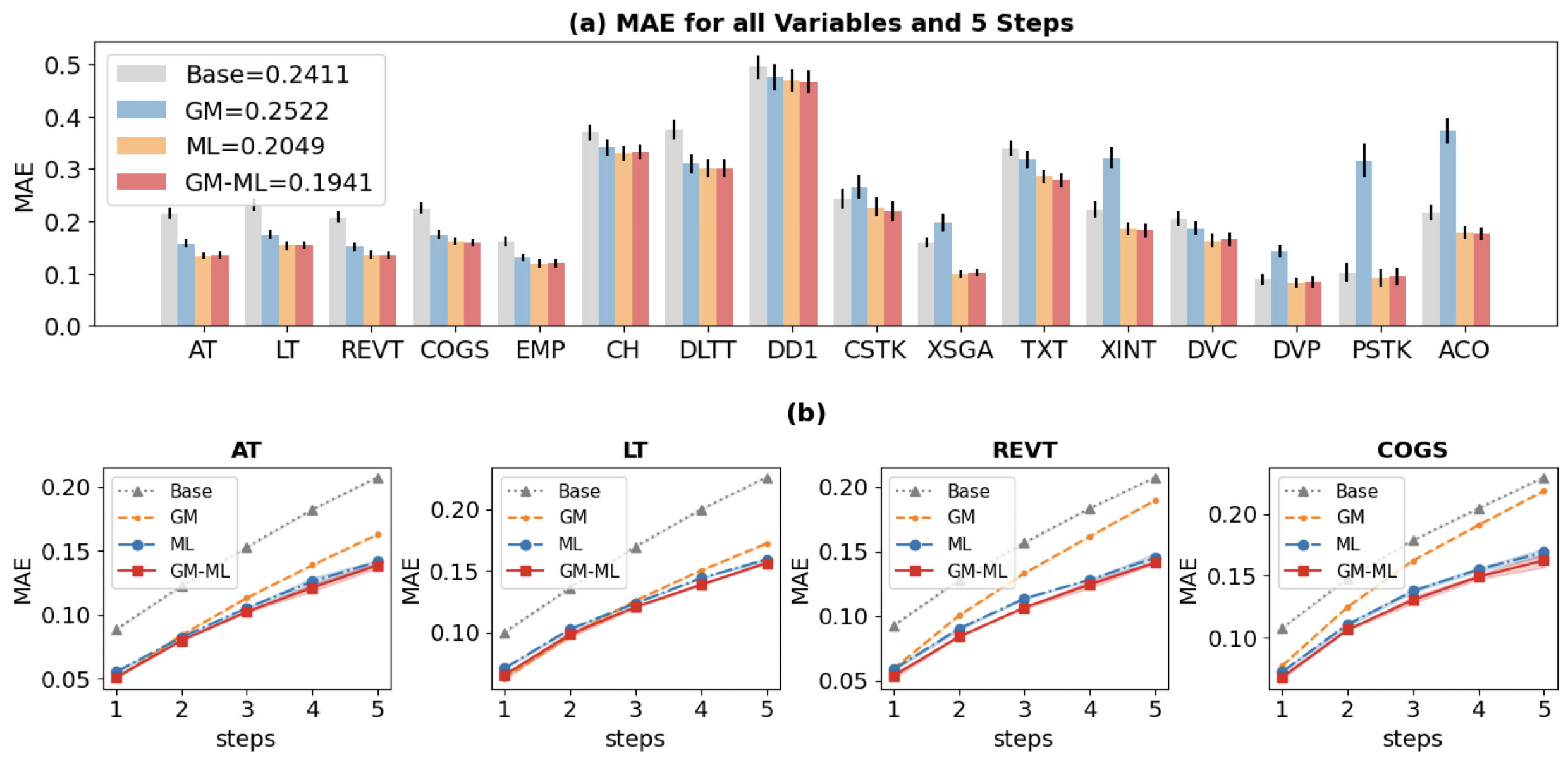}
    \caption{\textbf{Performance comparison of STIML and baseline models}. (a). Bar chart of the average MAE across 16 predicted variables for 5-step-ahead forecasting. The error bars indicate the standard deviation across repeated experimental runs. The legend reports the overall average MAE across all variables for each model. (b). The variation of prediction error with the prediction step for four core financial indicators, including Assets (AT), Liabilities (LT), Revenues(REVT) and Cost of Goods(COGS). The shadow areas indicate the standard deviation across repeated experimental runs. The corresponding step-wise error curves for all 16 target variables are provided in \ref{fig:fin_maes_si}.\iffalse (a) Comparison of average MAE for the 16 target predicted variables. The numbers in the legend are average MAE of all the variables. (b) Visualization of MAE for each prediction step. The whole variables could be seen in Figure \ref{fig:fin_maes_si}\fi}

    \label{fig:fin_maes}
\end{figure}

Figure \ref{fig:fin_maes} presents a comprehensive comparison between our model and the benchmark (MLP is used as the ML model here). {First, the extended GM outperforms the baselines in most cases (blue versus gray in Figure \ref{fig:fin_maes}(a)), indicating that it successfully captures part of the growth trend.} Reported results are averaged over three runs with different random seeds.
 Moreover, its performance is closely tied to the scaling relationship of the corresponding attribute: the better the scaling law fits (as reflected by the $R^2$ value of the power-law fit in Figure \ref{fig:scaling}), the better the extended GM performs.

{Second, results show that our hybrid model consistently achieves superior performance}. On average, it improves prediction accuracy by 23\% compared with GM and by 5\% compared with ML. Moreover, as shown in the figure, the hybrid model performs especially well when the GM itself provides a strong fit. We will return to this point in greater detail in the following discussion.

In Figure \ref{fig:fin_maes}(b) we visualize the MAE change with prediction steps and focus on the prediction results of four core aggregate indicators: assets (AT), revenues (REVT), cost of goods sold (COGS), and liabilities (LT). These four metrics capture distinct dimensions of company performance—namely, company size, revenue-generating capacity, cost efficiency, and leverage level. Many of the remaining indicators either serve as alternative measures of these dimensions (e.g., employee (EMP) count as an alternative proxy for company size), represent disaggregated components (e.g., long-term (DLTT) and short-term debt (DD1)), or reflect composite values (e.g., equity as the difference between assets and liabilities). The prediction results for all 16 financial indicators are provided in Figure \ref{fig:fin_maes_si}. In the following experiments, we adopt the same principle for visualization.

\subsection{Robustness: Alternative Machine Learning Backbones}\label{sec:robust}

To examine the robustness and generality of our framework, we conducted controlled comparisons across different ML architectures. For each architecture, we evaluated its performance within the hybrid framework and compared it to its purely data-driven counterpart. Table \ref{tab:model_performance} reports the MAE results for the four core indicators, and bold values indicate the lower error within each pairwise comparison.} 


\begin{table*}[t]
\centering
\small
\resizebox{\textwidth}{!}{%
\begin{tabular}{|c|c|c|c|c|c|c|}
\hline
\multirow{2}{*}{\textbf{ Model }}&\multirow{2}{*}{\textbf{ MAE}} & \multicolumn{5}{c|}{\textbf{Separated MAE}}\\
\hhline{|~~-----|}
&&Steps &AT&LT&REVT&COGS\\
\hline

\multirow{4}{*}{Baseline} & \multirow{4}{*}{0.1318} 
  & 1& 0.0885 & 0.1003 & 0.0925 &0.1077\\
&&2 & 0.1226 & 0.1361 & 0.1279 &0.1478\\
&&3 &0.1529 & 0.1696 & 0.1567 &0.1789\\
&&mean & 0.1213 & 0.1353 & 0.1257 &0.1448\\

\hline
\hline

\multirow{4}{*}{GM} & \multirow{4}{*}{0.0992} 
  & 1&0.0617 & 0.0650 & 0.0845 &0.0761\\
&&2 & 0.0970 & 0.0981 & 0.1229 &0.1150\\
&&3 & 0.1218 & 0.1196 & 0.1502 &0.1398\\
&&mean & 0.0935 & 0.0942 & 0.1192 &0.1103\\

\hline
\hline
\multirow{4}{*}{Random Forest}& \multirow{4}{*}{0.0977}  
&1 &  0.0696(0.0001) & 0.0854(0.0004) & 0.0764(0.0000) &0.1071(0.0003)\\
&& 2 & 0.0846(0.0003) & 0.1015(0.0010) & 0.0907(0.0005) &0.1190(0.0015)\\
&& 3 & 0.0956(0.0007) & 0.1122(0.0006) & 0.1014(0.0000) &0.1285(0.0008)\\
\hhline{|~~-----|}
&&  mean & 0.0833 & 0.0997 & 0.0895 &0.1182\\
\hline

\multirow{4}{*}{GM-Random Forest}& \multirow{4}{*} {\textbf{0.0793}} 
& 1 & \textbf{0.0474}(0.0005) & \textbf{0.0615}(0.0005) & \textbf{0.0516}(0.0005) &\textbf{0.0649}(0.0008)\\
&& 2 & \textbf{0.0718}(0.0004) & \textbf{0.0888}(0.0009) & \textbf{0.0767}(0.0009) &\textbf{0.0935}(0.0012)\\
&&  3 & \textbf{0.0875}(0.0009) & \textbf{0.1042}(0.0014) & \textbf{0.0925}(0.0009) &\textbf{0.1114}(0.0010)\\
\hhline{|~~-----|}
&&  mean & \textbf{0.0689} & \textbf{0.0848} & \textbf{0.0736} &\textbf{0.0899}\\

\hline
\hline
\multirow{4}{*}{VARIMA}& \multirow{4}{*}{0.2071}  
&1 & 0.0553(0.0009) & 0.0712(0.0004) & 0.0586(0.0007) &0.0723(0.0008)\\
&& 2 & 0.0824(0.0001) & 0.1029(0.0009) & 0.0902(0.0012) &0.1106(0.0013)\\
&& 3 & 0.1054(0.0003) & 0.1238(0.0003) & 0.1135(0.0001) &0.1378(0.0013)\\
\hhline{|~~-----|}
&&  mean & 0.0810 & 0.0993 & 0.0874 &0.1069\\
\hline

\multirow{4}{*}{GM-VARIMA}& \multirow{4}{*}{\textbf{0.0889}}  
& 1 & \textbf{0.0510}(0.0003) & \textbf{0.0656}(0.0017) & \textbf{0.0538}(0.0022) &\textbf{0.0679}(0.0013)\\
&& 2 & \textbf{0.0799}(0.0001) & \textbf{0.0985}(0.0022) & \textbf{0.0840}(0.0001) &\textbf{0.1063}(0.0003)\\
&&  3 & \textbf{0.1024}(0.0004) & \textbf{0.1207}(0.0006) & \textbf{0.1066}(0.0006) &\textbf{0.1305}(0.0026)\\
\hhline{|~~-----|}
&&  mean &  \textbf{0.0778} & \textbf{0.0949} & \textbf{0.0814} &\textbf{0.1016}\\

\hline
\hline

\multirow{4}{*}{MLP}& \multirow{4}{*}{0.0708}  
& 1 &\textbf{0.0289}(0.0002) & \textbf{0.0352}(0.0003) & 0.0298(0.0003) &0.0391(0.0002)\\
&&  2 &   0.0463(0.0004) & \textbf{0.0539}(0.0001) & 0.0481(0.0006) &0.0590(0.0000)\\
&&  3 & \textbf{0.0607}(0.0005) & \textbf{0.0695}(0.0003) & 0.0623(0.0003) &0.0737(0.0003)\\
\hhline{|~~-----|}
&&  mean &  \textbf{0.0453} & \textbf{0.0529} & 0.0467 &0.0573\\
\hline

\multirow{4}{*}{GM-MLP} & \multirow{4}{*}{\textbf{0.0697}}
& 1 & 0.0290(0.0002) & 0.0359(0.0003) & \textbf{0.0290}(0.0006) &\textbf{0.0372}(0.0012)\\
&& 2 & 0.0463(0.0001) & 0.0545(0.0003) & \textbf{0.0468}(0.0002) &\textbf{0.0573}(0.0005)\\
&& 3 & 0.0610(0.0002) & 0.0698(0.0003) & \textbf{0.0612}(0.0000) &\textbf{0.0726}(0.0003)\\
\hhline{|~~-----|}
&& mean & 0.0454 & 0.0534 & \textbf{0.0457} &\textbf{0.0557}\\

\hline
\hline

 \multirow{4}{*}{iTransformer} & \multirow{4}{*}{0.0711} 
& 1 & 0.0460(0.0169) & 0.0545(0.0138) & \textbf{0.0454}(0.0204) &\textbf{0.0481}(0.0201)\\
&& 2 & \textbf{0.0701}(0.0251) & \textbf{0.0783}(0.0201) & 0.0707(0.0281) &0.0724(0.0260)\\
&& 3 & 0.0898(0.0280) & 0.0973(0.0219) & 0.0905(0.0308) &0.0905(0.0262)\\
\hhline{|~~-----|}
 && mean& 0.0687 & 0.0767 & 0.0688 &\textbf{0.0703}\\
\hline

\multirow{4}{*}{GM-iTransformer}& \multirow{4}{*}{\textbf{0.0706}} 
& 1 & \textbf{0.0450}(0.0177) & \textbf{0.0533}(0.0155) & 0.0456(0.0202) &0.0488(0.0200)\\
&&2 & 0.0707(0.0258) & 0.0785(0.0204) & \textbf{0.0690}(0.0262) &\textbf{0.0721}(0.0249)\\
&&3 & \textbf{0.0891}(0.0289) & \textbf{0.0967}(0.0226) & \textbf{0.0885}(0.0303) &\textbf{0.0903}(0.0266)\\
\hhline{|~~-----|}
&&mean & \textbf{0.0683} & \textbf{0.0762} & \textbf{0.0677} &0.0704\\
\hline

\end{tabular}%
}
\caption{\textbf{ MAE of different models}. The table presents a comparison of prediction errors between different time-series forecasting models and their corresponding GM-enhanced versions. Bold values indicate the lower error within each pairwise comparison.\iffalse Bold values highlight the best-performing model within each comparison group. \fi}
\label{tab:model_performance}
\end{table*}

{Overall, the hybrid models outperform in most cases. However, increased architectural complexity can reduce generalization and increase the risk of overfitting when modeling fluctuations.} From table \ref{tab:model_performance}, although neural networks (NN) have demonstrated strong performance across a wide range of tasks, they do not show a clear advantage over traditional ML models in the context of company growth prediction on average. Moreover, between the two NN models, even iTransformer—considered the state-of-the-art for time series forecasting—does not outperform the simpler MLP when controlling for parameter size. 

\subsection{When Does Each Component Contribute? Subgroup Analyses}\label{sec:variables}

In this section, we analyze the contributions of different components  by comparing their accuracy in different groups. {Unless otherwise specified, we report the results of the best-performing RF model as shown in Table \ref{tab:model_performance}.}

\subsubsection{Size}

\begin{figure}
    \centering
    \includegraphics[width=1\linewidth]{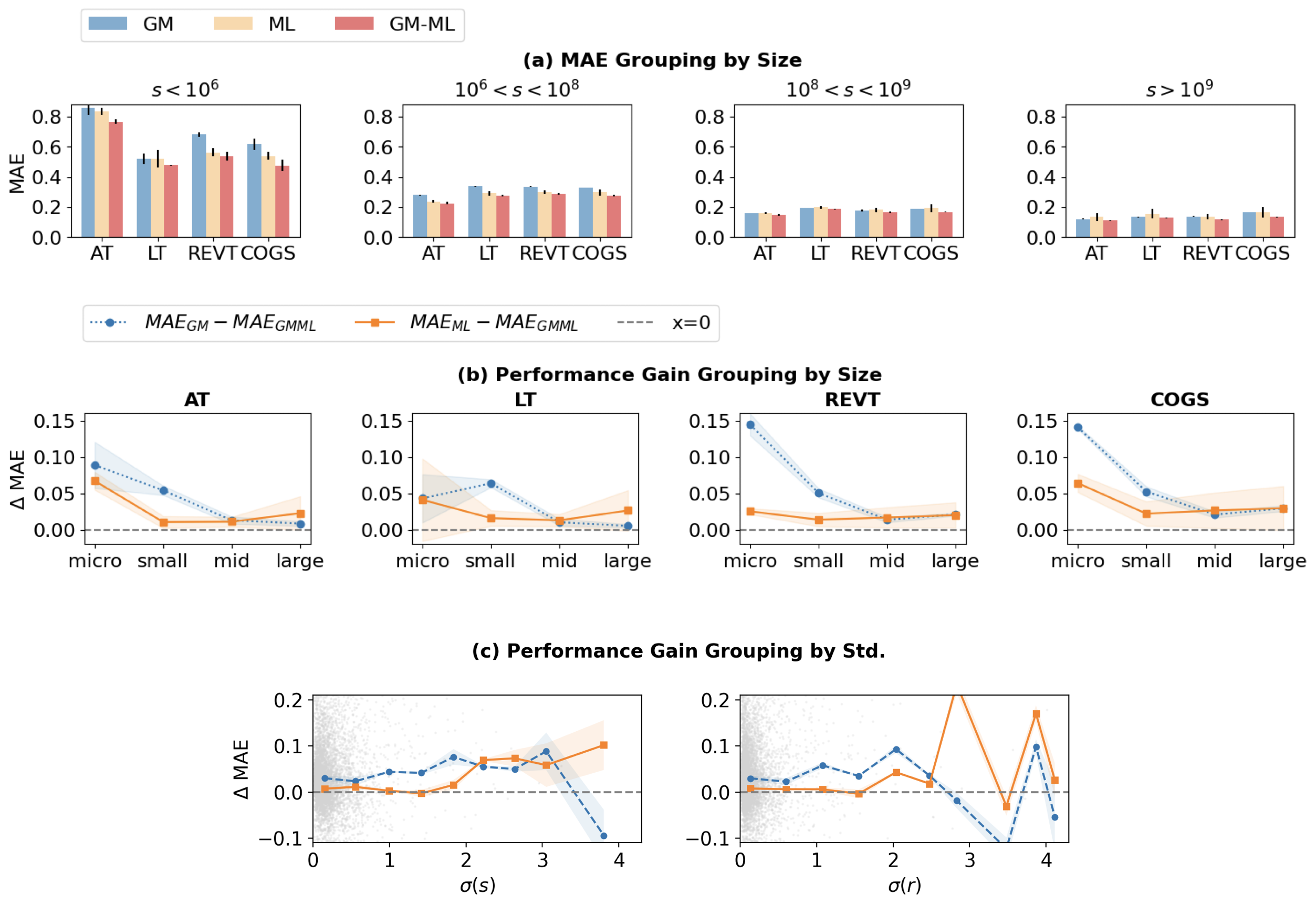}
    \caption{\textbf{Comparing model performance across different grouping strategies.}(a) Comparison of the MAE across different size groups of companies for different models. Each group corresponds to a range of average company assets (in USD), with categories defined as follows: micro $\in (0,10^6]$, small $\in [10^6, 10^8]$, mid $\in [10^8, 10^9]$, and large $\in [10^9,\infty]$. Different subplots represent different size groups. Each subplot displays the average MAE for three models—GM, ML, and GM-ML—on four core variables. The bars are color-coded as follows: blue for GM, yellow for ML, and red for GM-ML. (b) and (c) illustrates how the difference in average MAE between models varies across different subgroups. The blue dashed line represents the average MAE of GM minus that of GM-ML, while the yellow solid line represents the average MAE of ML minus GM-ML. The shaded areas denote the standard deviation range across multiple experimental runs. {Panel (b) groups companies by size, while Panel (c) groups them by standard deviation. $\sigma(s)$ is the std. of size and $\sigma(r)$ is the std. of growth rate. The gray dots in the background of panel (c) are the original company data.}}
    \label{fig:size}
\end{figure}

{Our initial findings highlight company size as a key explanatory factor for predictive performance (Figure \ref{fig:size})}. { Across all three models, there is a strong positive correlation between company size and prediction accuracy(Note the changing y-axis range in Figure \ref{fig:size}(a)). As size increases, the MAE of all three models significantly decreases. }

In general, larger companies yield more accurate forecasts than smaller ones, consistent with growth dynamics that are more strongly governed by systematic trends and exhibit more stable fluctuation structure. 
This can be attributed to several characteristics of large companies: they typically follow more established growth patterns, benefit from consistent revenue streams, operate with mature business models, and are less affected by market volatility. These features lead to more predictable behaviors that are easier for models to capture.

\subsubsection{Performance Gain}



To quantify the relative contribution of different growth components, we define \emph{performance gain} as the difference in average MAE between models.
Since GM-ML explicitly separates trend and fluctuation, this allows us to assess their marginal contributions by comparing the MAE of GM and GM-ML. 

Our results show that predictability is primarily driven by the trend component, particularly for large companies. As firm size increases, the prediction error of GM decreases sharply, while the performance gap between GM and GM-ML narrows(Blue dashed line in Figure \ref{fig:size}(b)). This pattern indicates that systematic trends account for an increasing fraction of predictive structure in large firms, whereas fluctuations contribute relatively less additional information.

However, clear heterogeneity emerges across size groups. For small companies, both trend and fluctuation components contribute substantially to predictability, giving the hybrid GM-ML model a distinct advantage. For large companies, growth trajectories are sufficiently trend-dominated that standalone GM or ML models already achieve strong performance, reducing the marginal benefit of hybrid modeling.

A natural conjecture is that our model, by explicitly separating trend and fluctuation, gains an advantage in high-volatility settings—since trend is relatively easier to predict, while volatility poses greater challenges. Small companies typically exhibit higher volatility, making this scenario especially relevant. 
To further test this mechanism, we examine the relationship between volatility and predictive performance gain. As shown in Fig.~\ref{fig:size}(c), performance gain increases systematically with volatility.
This finding confirms that structured fluctuations contain learnable predictive information, and that explicitly disentangling trend and fluctuation components provides the greatest benefit in high-volatility regimes.

{This suggests that volatility can be informative. For smaller or fast-evolving companies, managers should view fluctuations not as noise but as signals of adaptation or innovation cycles. By contrast, stable large companies may focus on strengthening underlying trends through efficiency gains and long-term resource planning.}

Together, these results support the central hypothesis of the STIML framework: company growth consists of two predictable components—systematic trends and structured fluctuations—whose relative importance varies across firms.

\subsubsection{Error Asymmetry}

\begin{figure}[h]
    \centering
    \includegraphics[width=1\textwidth]{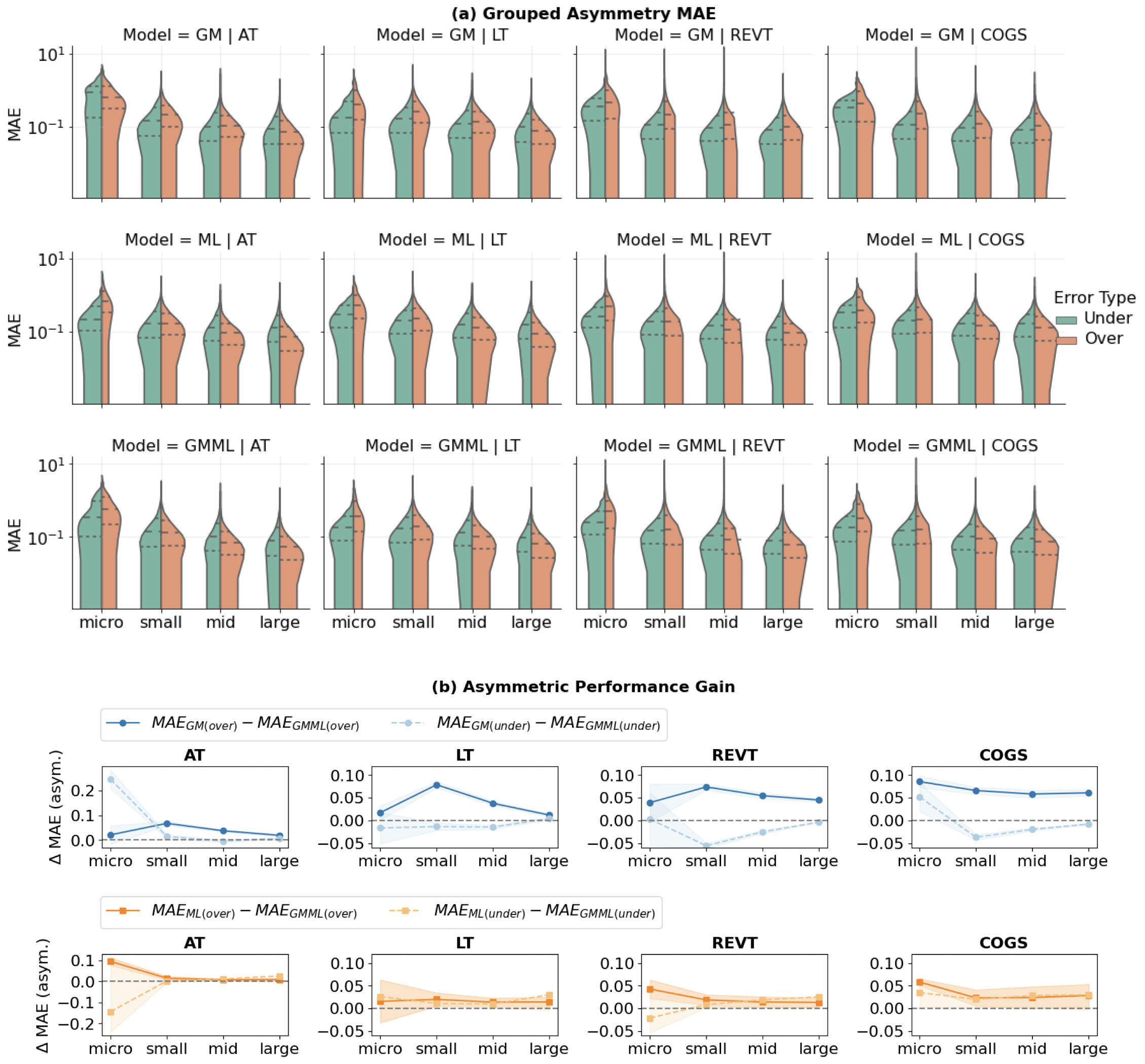}
    \caption{\textbf{Error asymmetry analysis.} (a) MAE distributions by error sign (over- vs. underestimation) for the four core financial indicators under GM (first row), ML (second row), and GM-ML (third row). Green denotes the underestimation MAEs, and orange denotes the overestimation MAEs. (b) Asymmetric performance gains, with gain curves decomposed into over- and underestimation components.}
    \label{fig:asymmery}
    
\end{figure}

In this part, we further decompose the prediction errors by their signs and examine whether the performance of each model is symmetric with respect to over- versus under-estimation. Figure \ref{fig:asymmery}(a) reports the MAE distributions conditioned on the error signs. Specifically, for each target variable and each model (GM, ML, and GM-ML), we split the test samples into two groups according to whether the model over-predicts ($\hat{Y}>Y$) or under-predicts ($\hat{Y}<Y$), and then visualize the MAE distribution within each group. {The results reveal a clear asymmetry in Figure \ref{fig:asymmery}. The error distributions of over- and under-estimation differ substantially across the four core indicators (AT, LT, REVT, COGS). }

Using AT as an illustrative example, we examine the latent patterns captured by GM-ML. Comparing the first column of Figure \ref{fig:asymmery}(a), {we observe a clear asymmetry for GM in the micro-size group: negative residuals are more prevalent, indicating that GM tends to systematically under-predict the growth of small companies\cite{zhang2021scaling}. In contrast, relative to ML, GM-ML provides a noticeably stronger correction for the under-estimation cases}. Note that this comparison is aggregated and should be interpreted as a coarse, distributional contrast, not as implying that the same firms appear in each subgroup.

Figure \ref{fig:asymmery}(b) further decomposes the performance gain curves in Figure \ref{fig:size}(b) into over- and under-estimation components. Specifically, we compute the MAE difference between each baseline model (GM or ML) and GM-ML separately within the over-estimation subset and within the under-estimation subset. {This decomposition clarifies that the improvement of GM-ML primarily comes from correcting systematic over-prediction} (Most of the solid lines in Figure \ref{fig:asymmery}(b) lie above the zero line (y=0), whereas the dashed lines fluctuate around zero with no clear structure). That is to say, the negative residuals contain more structured information. 

{This pattern is consistent with the view that, when the system operates under binding constraints or frictions, output is systematically depressed by identifiable mechanisms—yielding structured negative errors—whereas above-expectation gains are more often driven by idiosyncratic opportunities and unobserved factors, and thus exhibit weaker structure and higher noise.}

\subsection{Opening the Black Box: Interpretability analysis}\label{sec:interpret}

\subsubsection{Feature Representation}\label{sec:representation}

To understand how the neural network internally represents companies, we analyze the learned latent representations in the trained model. Specifically, we extract the high-dimensional hidden-layer representations passed from the encoder to the decoder, which provide a compact and informative embedding of each company. These latent representations capture key structural information learned by the model and reflect how companies are organized in the model’s internal feature space.
In this analysis, we use the latent representations from the GM-iTransformer model. To visualize the structure of this high-dimensional representation space, we apply principal component analysis (PCA) to project the latent vectors into a two-dimensional space. This projection allows us to examine the geometric organization of companies in the learned feature space.

\begin{figure}
    \centering
    \includegraphics[width=1\linewidth]{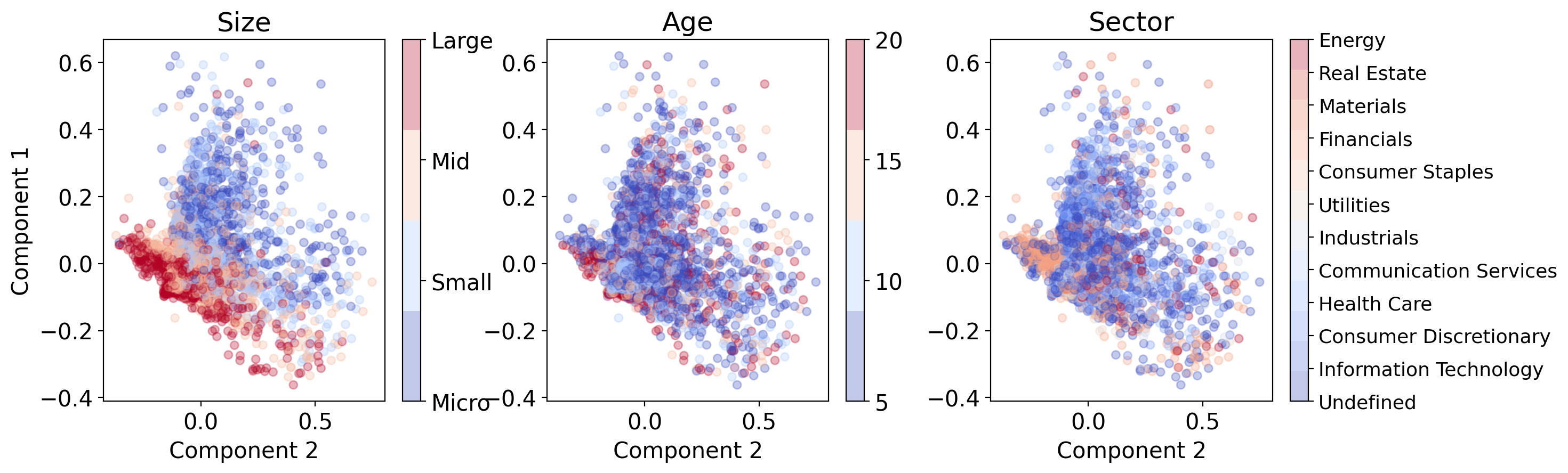}
    \caption{\textbf{Visualization of feature representations learned in GM-iTransformer model.} PCA visualization of neural-network hidden-layer feature representations, colored by firm size (left), age (middle), and sector (right).}
    \label{fig:representation}
\end{figure}

{Figure \ref{fig:representation} indicates that the feature representation space is strongly structured by company size. This relationship holds consistently across different size measures, including AT, REVT, or the EMP.} By contrast, other attributes, such as age or sector, exhibit weaker and less consistent organization in the latent space. {This finding is consistent with our earlier results, indicating that company size is the dominant factor governing both predictive performance and the internal representation learned by the model.} 

\subsubsection{Features Importance Decomposition}\label{sec:explainer}

\begin{figure}
    \centering
    \includegraphics[width=0.9\linewidth]{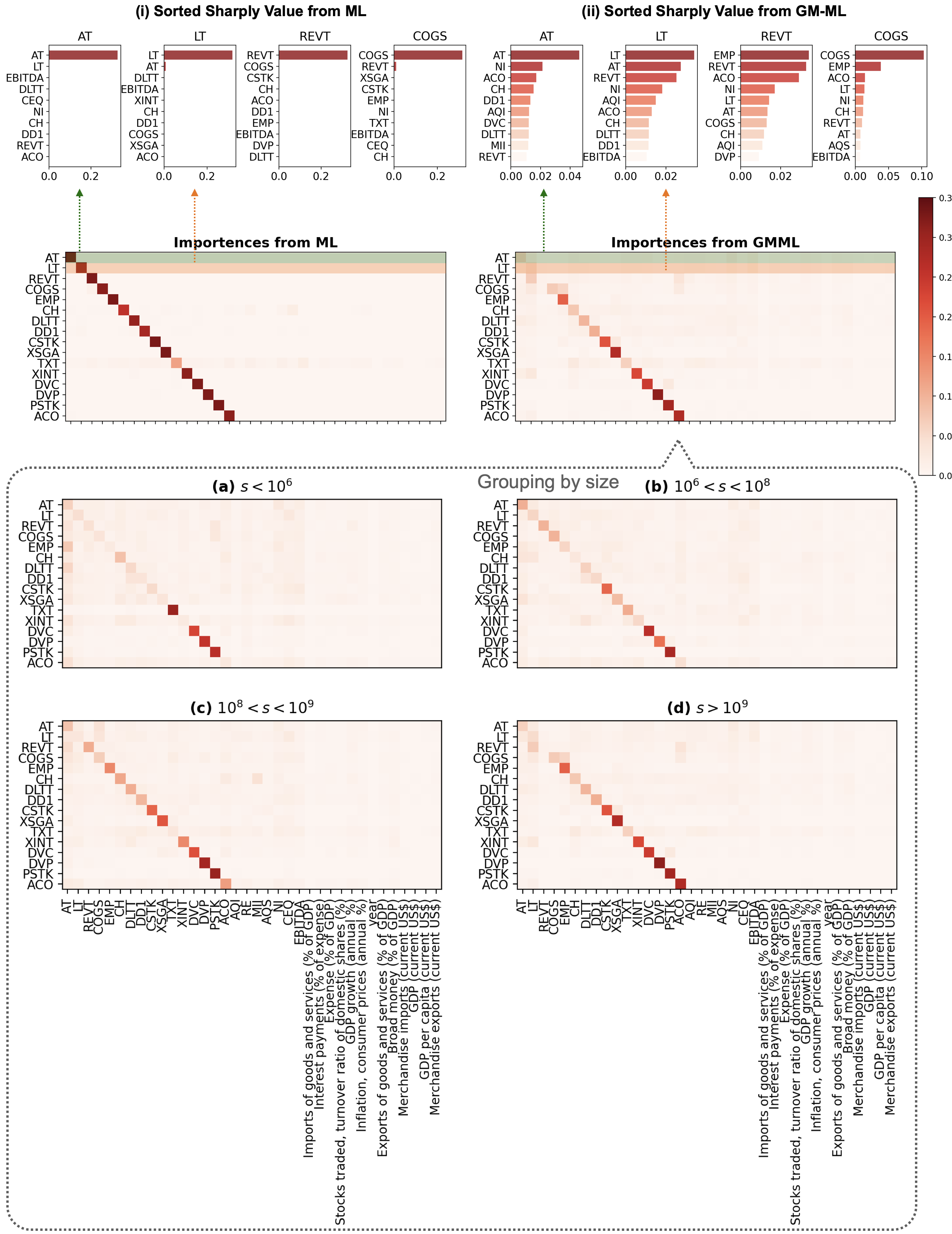}
    \caption{\textbf{Feature importance analysis via SHAP.} The middle heatmap reports mean absolute SHAP values, where columns correspond to input variables and rows correspond to prediction targets (darker colors indicate larger contributions). The top bar plots (i) and (ii) provide a zoomed-in view (top 10 features) for selected targets highlighted in the heatmap (e.g., AT and LT, indicated by the green and orange arrows, respectively). Panels (a)–(d) further report SHAP value patterns of the GM-ML model across firm-size groups.}
    \label{fig:shap}
\end{figure}


To gain a deeper understanding of the behavior and interpretability of our model, {and to examine the role of incorporating GM, we analyze feature importance using SHAP values\cite{lundberg2017unified}. }



Figure \ref{fig:shap} presents the feature importance for both the purely ML and GM-ML models. The heatmaps in the middle line show the contribution matrix of input variables (columns) to each prediction target (rows), where color intensity represents the mean absolute Shapley value aggregated over the temporal dimension.
Panels (i) and (ii) provide detailed views of the top 10 features for selected targets highlighted in the heatmap (e.g., AT and LT, marked by the green and orange arrows, respectively).

An interesting observation is that the ML model relies primarily on autocorrelation, with each variable’s own lagged values dominating its prediction. Contributions from other variables remain comparatively weak. In contrast, the GM-ML shows a clearer multivariate dependency structure that extends beyond autocorrelation.
This indicates that the GM-ML model is capable of leveraging richer inter-variable relationships to improve predictive performance. {Our robust test shows the ranking of feature importance is insensitive to the order of input features — shuffling the input variables does not substantially affect their importance ranking.}

Another interesting finding is that macroeconomic variables contributed relatively little to the prediction accuracy. This pattern is consistent across prediction targets and remains robust when companies are grouped by size, as shown in Fig.~\ref{fig:shap}(a–d). This suggests that company growth dynamics are primarily governed by internal structural factors rather than external macroeconomic fluctuations, at least in terms of predictive relevance.

Furthermore, the SHAP analysis reveals systematic differences across firm sizes. Smaller companies exhibit weaker autocorrelation and greater reliance on multivariate relationships, whereas larger companies are more strongly governed by their own historical trajectories. 

Together, these results confirmed that explicitly disentangling trend and fluctuation components enables the model to uncover structured predictive relationships that remain hidden in purely data-driven approaches.




\section{Discussion and Conclusion}\label{sec:conclusion}

{Company growth exhibits a structured coexistence of trends and heterogeneous fluctuations, a hallmark of complex adaptive systems.} This study proposes a hybrid framework, STIML, for forecasting company financial indicators. A central contribution of STIML is the explicit decomposition of growth dynamics into a mechanistic trend and residual fluctuations. By extending the scaling-based growth model (GM) to multiple financial variables to capture system-level average growth trends, and combining it with machine learning (ML) to model fluctuations, STIML delivers high predictive accuracy while improving interpretability. Extensive experiments on the Compustat dataset show that STIML significantly outperforms both purely GM and purely ML methods. On average, we find that our approach improves prediction accuracy by 23\% compared with GM and by 5\% compared with ML, highlighting the benefits of integrating growth mechanism with flexible machine learning models.

This decomposition further reveals clear heterogeneity and regime-like behavior across companies. We found that the GM component excels in capturing average trends, particularly for large, stable companies with low volatility, while the ML component is more effective for smaller, high-risk firms where fluctuations dominate. This complementarity highlights the value of explicitly separating trend and fluctuation dynamics in modeling company growth. 

Feature importance analysis further clarifies the source of these improvements.  Pure ML models rely largely on autocorrelation, whereas STIML captures cross-variable interactions beyond simple persistence.  {Interestingly, we found macroeconomic variables contribute little on average, a result robust across company sizes. This suggests that company growth is largely governed by internal dynamics, and that most growth dynamics seem to be endogenous to companies as opposed to following macroscopic trends or exogenous conditions. Importantly, this result should not be interpreted as macroeconomic factors being irrelevant. Rather, at the firm level and in an average predictive sense, macro-level effects tend to cancel out across heterogeneous firms, leaving internal dynamics as the dominant source of predictability.} 


{At the same time, as a hybrid model, STIML reveals complementary limitations of both ML and GM. Compared with pure ML approaches, STIML captures predictable structure that is not fully exploited by standalone ML. This is reflected in consistent, systematic gains in predictive accuracy across firms of all sizes, not just within a specific size regime. This suggests that trend and fluctuation components in company growth time series are statistically separable, and that explicitly modeling trend information reveals predictable structure that standalone ML models do not fully exploit.}

{Compared with GM, STIML’s performance gains suggest that deviations from scaling laws are not mere noise but contain predictive signals. STIML exhibits systematic error correction in the negative-residual regime. This indicates that the residual component in GM—particularly negative deviations—contains structural information relevant for prediction that is not fully captured by the original GM framework.}
{In physical systems, scaling relations are typically used to capture scale-invariant behavior, where positive and negative fluctuations around the mean are often symmetric, as in the case of magnetization in the Ising model. Economic systems, however, are fundamentally different. Asymmetries introduced by risk aversion, bankruptcy thresholds, liquidation mechanisms, and policy interventions lead to directional biases that have no direct analogue in natural systems. While scaling laws may remain effective for capturing long-term average trends, they may discard important asymmetric fluctuation structures. Our results provide preliminary evidence that negative residuals in company growth contain structured information that is systematically ignored by symmetric scaling-based growth models, pointing to the need for future mechanistic models that explicitly account for asymmetric fluctuations.}

{In conclusion, we present STIML in this work, a scaling-theory-informed machine learning framework for company growth prediction. Our results offer a complexity-oriented view of company growth, suggesting it is neither purely random nor fully determined by universal laws. Instead, it reflects the interaction between system-level regularities and firm-specific fluctuations. The relative importance of these components depends on size and volatility, leading to distinct regimes of predictability across companies}

{More broadly, this work illustrates how explicitly separating trend and fluctuation can serve as a general modeling principle for complex systems. Although our analysis focuses on company growth, the proposed framework is not limited to corporate finance. Many economic and social systems exhibit universal scaling relations together with strong individual heterogeneity. Extending scaling-theory-informed learning to such systems may provide a systematic way to bridge mechanism-based models and modern data-driven methods.}


\bibliographystyle{unsrt}
\bibliography{cas-refs}

\section*{Data availability}

The data that support the findings of this study are available from the Compustat database.

\section*{Code availability}

The underlying code for this study (and training/validation datasets) could be available to qualified researchers on reasonable request from the corresponding author.

\section*{Acknowledgments}
We would like to express our gratitude to Swarma Club for their support towards our paper, as well as the constructive discussions contributed by Zhang Zhang, Mingze Qi and Lei Dong.

\clearpage

\setcounter{section}{0}
\setcounter{figure}{0}
\setcounter{table}{0}
\setcounter{equation}{0} 
\renewcommand{\thefigure}{S\arabic{figure}}
\renewcommand{\thetable}{S\arabic{table}}
\renewcommand{\theequation}{S\arabic{equation}}

\renewcommand{\thesection}{S\arabic{section}}
\renewcommand{\thesubsection}{S\arabic{section}.\arabic{subsection}}
\renewcommand{\thesubsubsection}{S\arabic{section}.\arabic{subsection}.\arabic{subsubsection}}

\section*{Supplementary}


\begin{figure}
    \centering
    \includegraphics[width=1\linewidth]{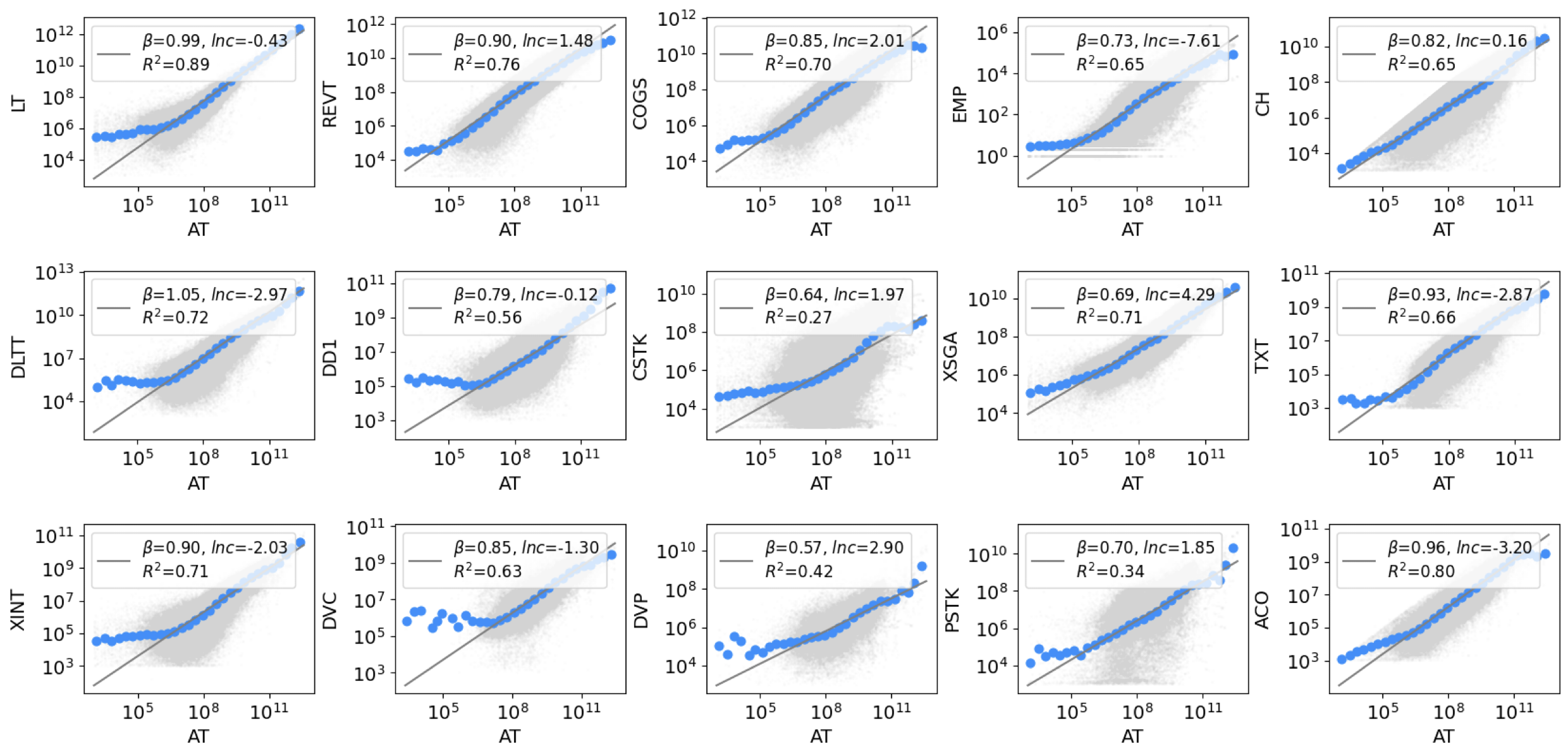}
    \caption{We have 23 financial indicators in total (See Table \ref{tab:finance} in SI). Among them, 15 indicators are positive and exhibit scaling relationships with assets. Therefore, at most 16 predictors can be used, including assets itself.}
    \label{fig:scaling}
\end{figure}

\section{Data Preprocessing}\label{sec:pre_progress_si}

We screened and preprocessed the raw data as follows:
\begin{enumerate}
    \item \textbf{Feature Selection}: First, we removed the features whose missing values were greater than 50\%. 
    \item \textbf{Noise reduction}: We filtered out companies with a time series of less than 2 years. Companies that appear in the dataset for only 1 or 2 years can neither serve as training samples nor be predictable. And then, we also filtered out the years in which abnormal data appear in financial statements, such as when liabilities, income costs, etc., are negative or zero. This type of data is considered anomalous and needs to be eliminated.
    \item \textbf{Missing value imputation}:
    There are some years with missing data in the dataset, and the previous step of anomaly handling may also leave spots where data are missing. In this paper's experiments, if the missing value is in the middle, we used the average of the time series data before and after to replace it; if the missing value is at the endpoints, we used the previous or subsequent value to replace it.
    \item \textbf{Inflation adjustment}:
    We applied inflation adjustments to all the monetary data across different years, with 2019 serving as the base year. The adjustments were made via the year-average inflation consumer price (ICP).
    \item \textbf{Logarithm transformation}:
    We took the natural logarithm of all monetary data for scaling. For some indicators that may have negative values (such as profit, cash flow, etc., data from the cash flow statement) and some values that occasionally occur as 1 (such as the number of employees), we have developed a linear-log method in practice, as shown in Equation \ref{eq:linear-log}. 
   
    \begin{equation}
    f(x) = \delta(x) \ln (|x| + 1), {x \neq 0} 
    \label{eq:linear-log}
    \end{equation}
$\delta(x)$ represents the sign of $x$, indicating whether it is positive or negative. This method can achieve the same effect as taking the logarithm: not only can it scale negative values similarly, but it can also be used within the 0-1 interval.
    
\end{enumerate}

\begin{table}[h!]
  \centering
    \caption{Financial Indicators}
    \label{tab:finance}
    \begin{tabular}{ l c c c } 
    \hline
      \textbf{Indicators} &\textbf{Description}& \textbf{Minimum} & \textbf{Maximum}\\
    \hline
      AT & Assets& \num{1.0240e6}& \num{3.7712e12}\\
      LT &Liabilities &\num{1.0460e3} & \num{3.7741e12}\\
      REVT & Revenue & \num{7.6300e2} &  \num{5.1479e11}\\
      COGS &Cost of Goods Sold & 0 & \num{3.9934e11}\\
      EMP & Employee &1 & \num{2.3000e6}\\
      CH & Cash& 0 & \num{1.6903e11} \\
      DLTT & Long-Term Debt & 0&\num{3.5579e12} \\
      DD1 & Long-Term Debt Due in One Year &0&\num{1.9923e11}\\
      CSTK & Common/Ordinary Stock (Capital)& 0 &\num{7.1649e10}\\
      XSGA & Selling, General and Administrative Expenses& 0& \num{1.0729e11}\\
      TXT & Income Taxes & 0 & \num{4.8919e10}\\
     XINT & Interest and Related Expenses& 0& \num{1.6136e11}\\
     DVC & Dividends Common/Ordinary & 0&\num{4.9146e10}\\
     DVP & Dividends - Preferred/Preference & 0& \num{9.3423e+10}\\
    PSTK & Preferred/Preference Stock (Capital) & 0 & \num{1.5213e+11}\\
    ACO & Current Assets - Other&0&\num{6.5765e10}\\
    AQI & Acquisitions - Income Contribution &\num{-1.4806e+08} & \num{6.5765e+10}\\
    AQS & Acquisitions - Sales Contribution &\num{-6.1114e+09} & \num{9.4562e+11}\\
    MII & Minority Interest (Income Account) &\num{-1.4352e+10} & \num{1.3179e+10}\\
    CEQ &Common/Ordinary Equity&\num{-1.5700e11} & \num{3.6440e11}\\
    EBIDTA & Earnings Before Interest & \num{-9.2932e10}&\num{1.4612e11}\\
    NI & Net Income (Loss) & \num{-1.3997e11} &\num{1.2225e11} \\
    RE &Return Earnings& \num{-1.4926e11}& \num{4.1977e11}\\
     \hline
    \end{tabular}
\end{table}

\begin{figure}
    \centering
    \includegraphics[width=1\linewidth]{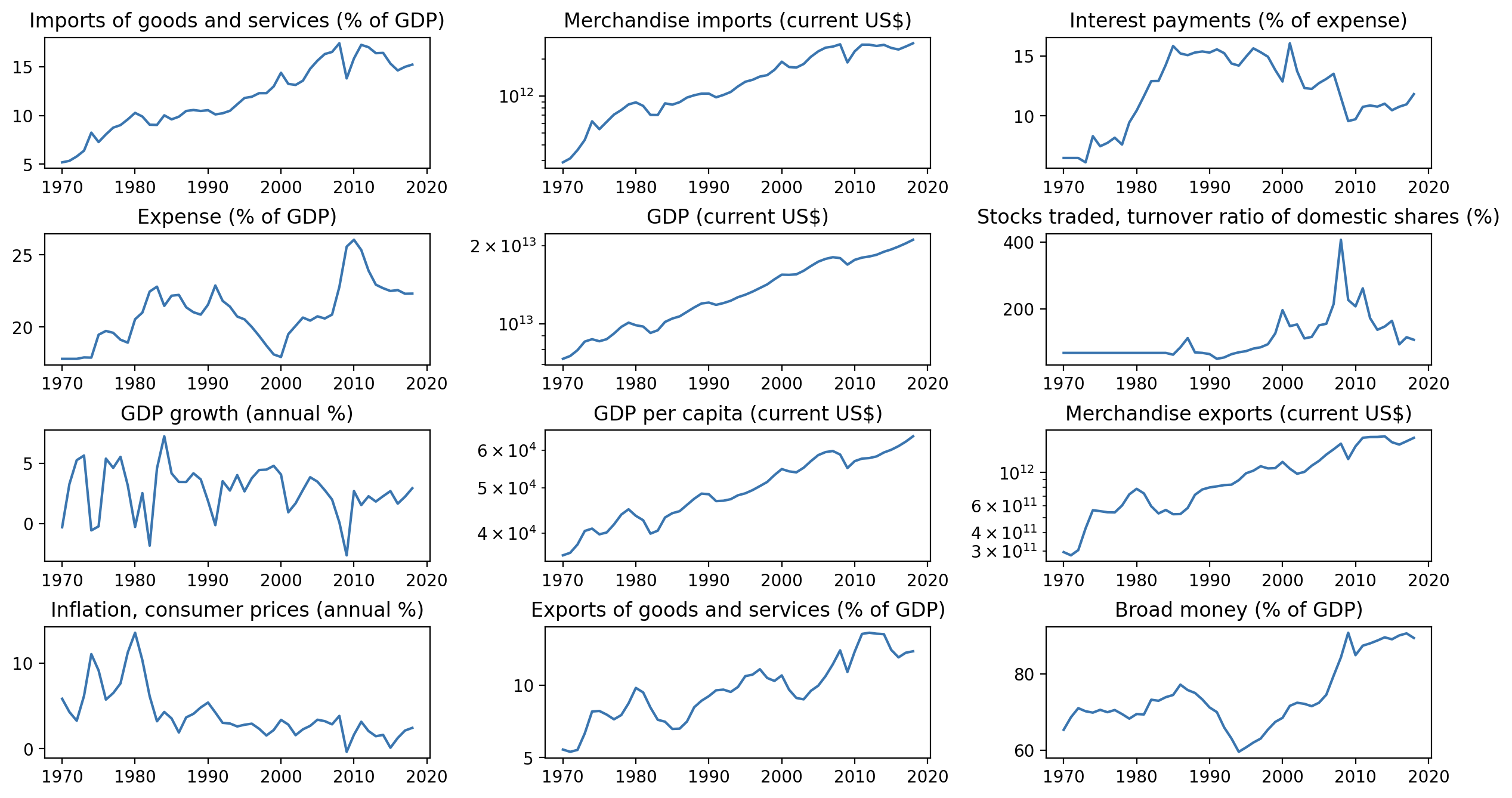}
    \caption{12 Macroeconomic Indicators from 1970 to 2019.}
    \label{fig:macro_index}
\end{figure}

\subsection{Sample split: Training, validation, and test data}

\begin{figure}
    \centering
    \includegraphics[width=1\linewidth]{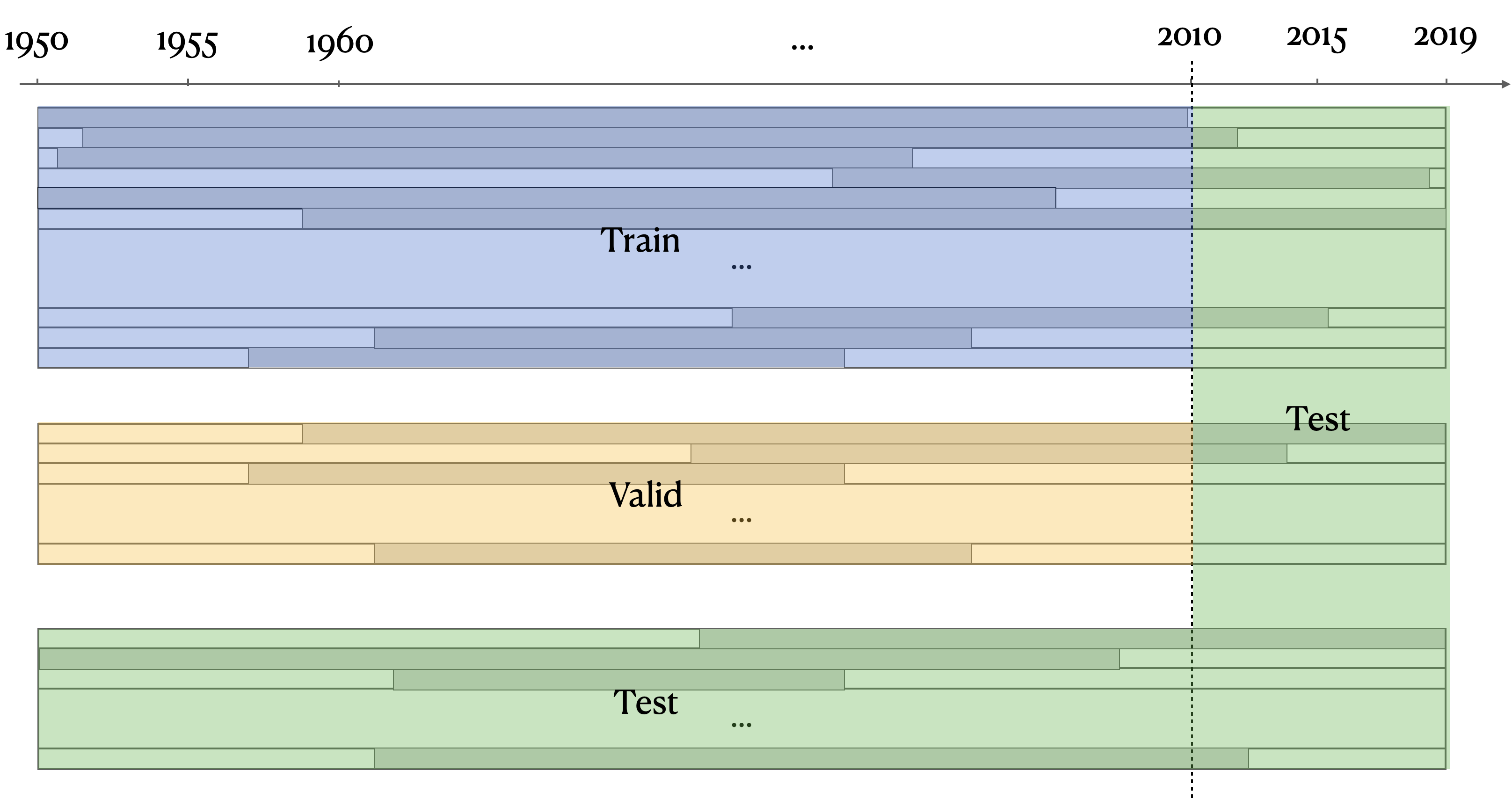}
    \caption{\textbf{Data setting}. Our dataset spans from 1950 to 2019. In the figure, each gray row represents a company, and the solid-colored segment indicates the years during which the company was active. We first partitioned the data into periods before and after 2010 (dash line). Companies established before 2010 were further divided into a training set (blue), validation set (yellow), and test set (green). In addition, all observations from 2010 onward were included in the test set.\iffalse Our dataset spans from 1950 to 2019. In the figure, each gray column represents a company, with the solid color indicating the survival years of that company. The age distribution is shown in figure (b). We initially partitioned the data into periods before and after 2010. For companies established before 2010, we further split them into a training set (blue), validation set (yellow), and test set (blue). All data from 2010 onward are also included in the test set.\fi}
    \label{fig:dataseting}
\end{figure}

We first split the data between post-2010 and pre-2010 and then divided the pre-2010 observations into training, validation, and testing sets at a 6:2:2 ratio according to the number of companies. That is, over 13,000 companies constitute in the training set, over 5,000 companies constitute the validation set, and over 5,000 companies constitute the test set. All the post-2010 data are also used for testing. Figure \ref{fig:dataseting} shows a schematic diagram of the dataset division. 
{The reason for this approach is that, unlike typical time series forecasting tasks, company-level financial statement data is characterized by having a large number of samples but relatively short time spans. Therefore, we choose to partition the dataset primarily based on the characteristics of the data. In most cases, different batches correspond to different companies rather than different time windows.}

The fitting of parameters for $GM(\beta_X, c_X)$ and the training of the machine learning are performed on the training set, which is shown the blue section of Figure \ref{fig:dataseting}.

\subsection{Choice of Hyperparameters}

{Length of historical time-series is set as $S=3$ and also the prediction length $T=3$ for the most cases.} The mean squared error (MSE) is chosen as the loss function, and the Adam optimizer is applied for gradient descent, with an adaptive learning rate and a weight decay of 0.005. Unless otherwise specified, the hidden size of the MLP in both the encoder and decoder is set to 32, and to 8 for iTransformer. All attention layers are configured with a single layer. 

\section{Detail of iTransformer Model}\label{sec:itrans_arch}

\begin{figure}
    \centering
    \includegraphics[width=1\linewidth]{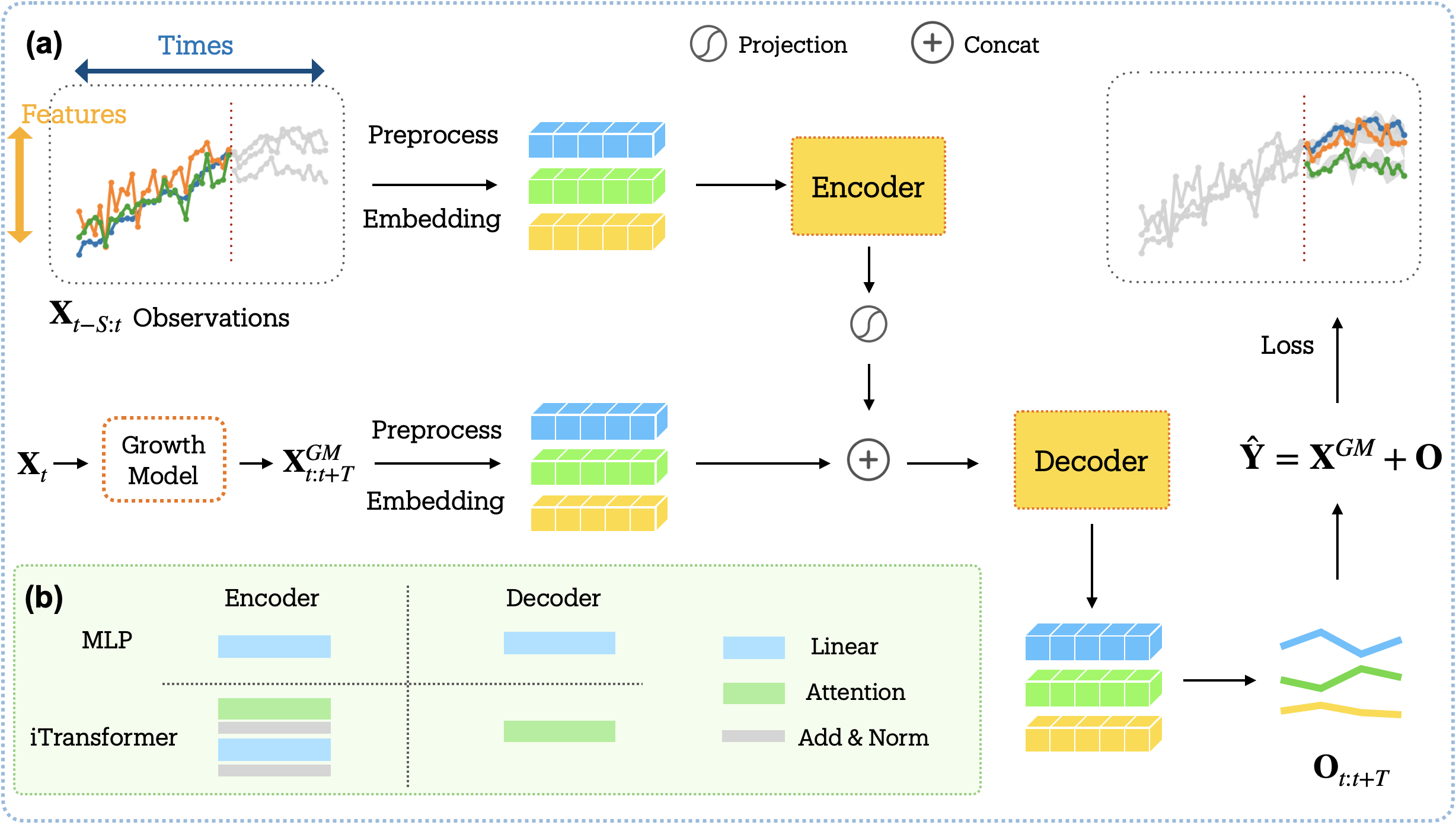}

    \caption{\textbf{Overall architecture of GM combined neural network.} (a) presents the complete framework from raw data processing to final prediction results; (b) illustrates the core components of both MLP and iTransformer architectures.}
    \label{fig:nn_framework}
\end{figure}

The input of encoder module is ${\mathbf{X}_{t-S:t,:}}$, including the {target variables ${\mathbf{X}_{t-S:t,:N}}$ and concatenates them with other potentially useful predictive variables, such as macroeconomic information $\mathbf{X}^{macro}_{t-S:t,:} \in \mathbb{R}^{S \times (N'-N)}$}. Ultimately, this part is modeled as follows:

\begin{equation}
    \begin{aligned}
        H_0 &= Embedding(\mathbf{X}_{t-S:t,:}), \\
       H_{l+1} &= Encoder(H_l),l=0,\dots,L-1 \\
       En_{l+1} &= Projection(H_L) \\
    \end{aligned}
\end{equation}

$Embedding$ and $Projection$ are both linear transformation here.  $\mathbf{X}$ is normalized before embedding. $H_l$ is the hidden state for the lth layer. For the MLP model, the encoder module consists of linear layers followed by activation functions. For the iTransformer architecture, we largely adhere to the settings described in \cite{liu2024itransformer}, including multivariate attention , 2 layer normalization, and feed-forward networks. Detailed architectures of both MLP and iTransformer are illustrated in Figure \ref{fig:nn_framework}. The final output denote as $En_{l+1}$ is passed into the decoder module.

In the decoder module, we predict future time steps from $t$ to $t+T$. The inputs include ${{\mathbf{X}_{t:t+T}}^{GM}}$, which is the prediction from the GM, macroeconomic information ${{\mathbf{X}_{t:t+T}}^{macro}}$ and,  of course,  the output from the encoder modules  $En_{l+1}$. 

\begin{equation}
    \begin{aligned}
    \mathbf{O} = Decoder({\mathbf{X}^{GM}}_{t:t+T} \bigoplus {\mathbf{X}^{macro}}_{t:t+T}, En_{l+1})
     \end{aligned}
\end{equation}

$\bigoplus$ is the concatenating operation. {For the MLP model, the decoder module similarly comprises linear layers with activation functions. Regarding the iTransformer architecture, since the original model lacks this component, we introduce an additional attention module to compute the attention matrix between extra input and $En_{l+1}$. Detailed architectures are illustrated in Figure \ref{fig:nn_framework}. After reshaping, the output of decoder module is $\mathbf{O} =\{\mathbf{o}_1, \dots, \mathbf{o}_T\}\in \mathbb{R}^{T \times N}$ } after adding back the output of GM  $\mathbf{X} ^ {GM}_{t:t+T}$, we can obtain the final prediction $\mathbf{\hat{Y}}_{t:t+T} = \mathbf{O}_{t:t+T} + \mathbf{X}^{GM}_{t:t+T}$. 

The loss function here is the mean square error (MSE), and it optimizes the difference between the prediction $ \mathbf{\hat{Y}} $ and the actual data $\mathbf{Y}$ in log space.

\begin{equation}
    Loss = \frac{1}{T}\Sigma_j^T||\ln Y_j - \ln\hat{Y_j}||^2
\end{equation}

\begin{figure}
    \centering
    \includegraphics[width=1\linewidth]{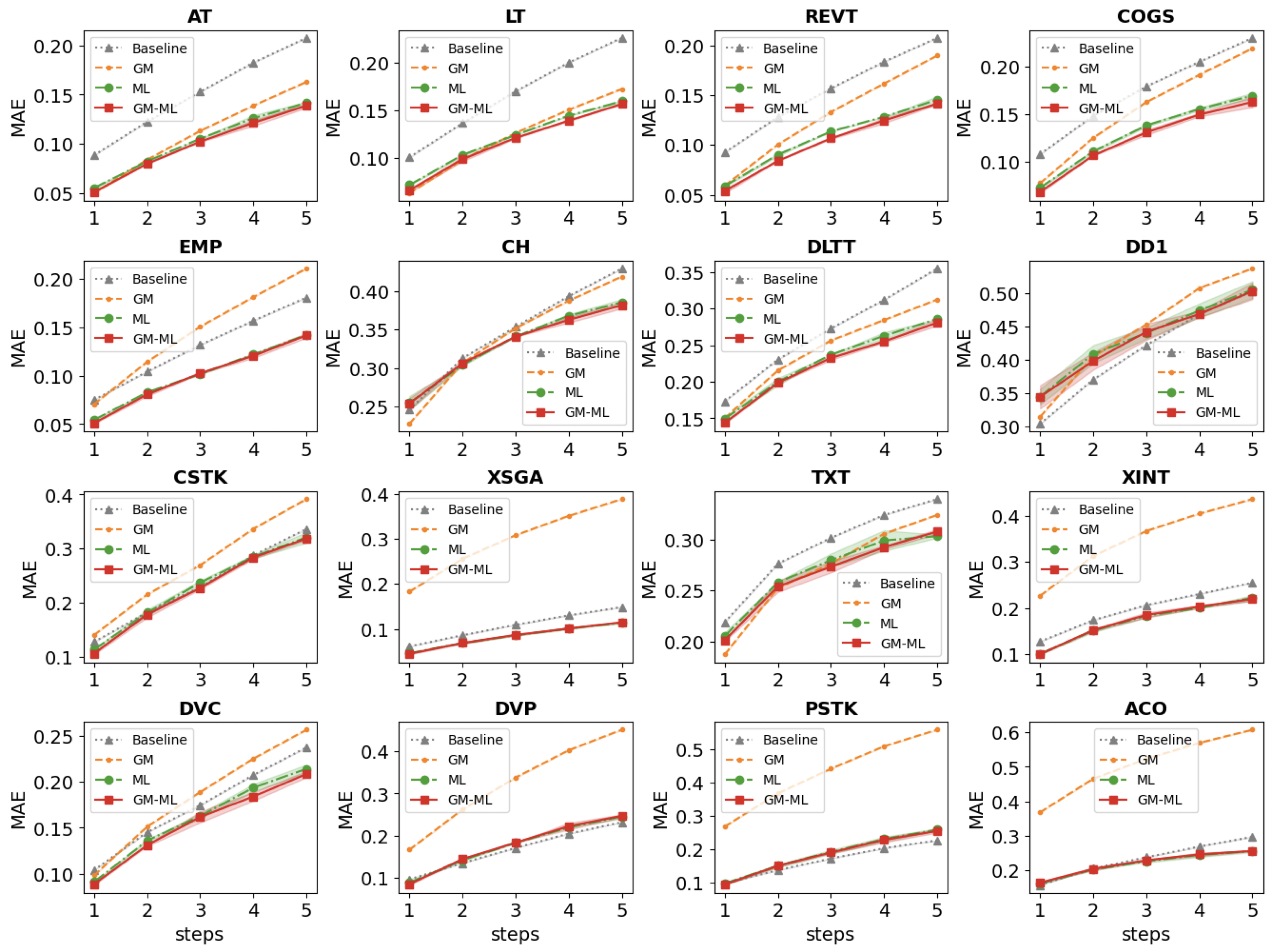}
    \caption{5-step-ahead MAE comparison across models for all predicted variables}
 
    \label{fig:fin_maes_si}
\end{figure}

\begin{figure}
    \centering
    \includegraphics[width=1\linewidth]{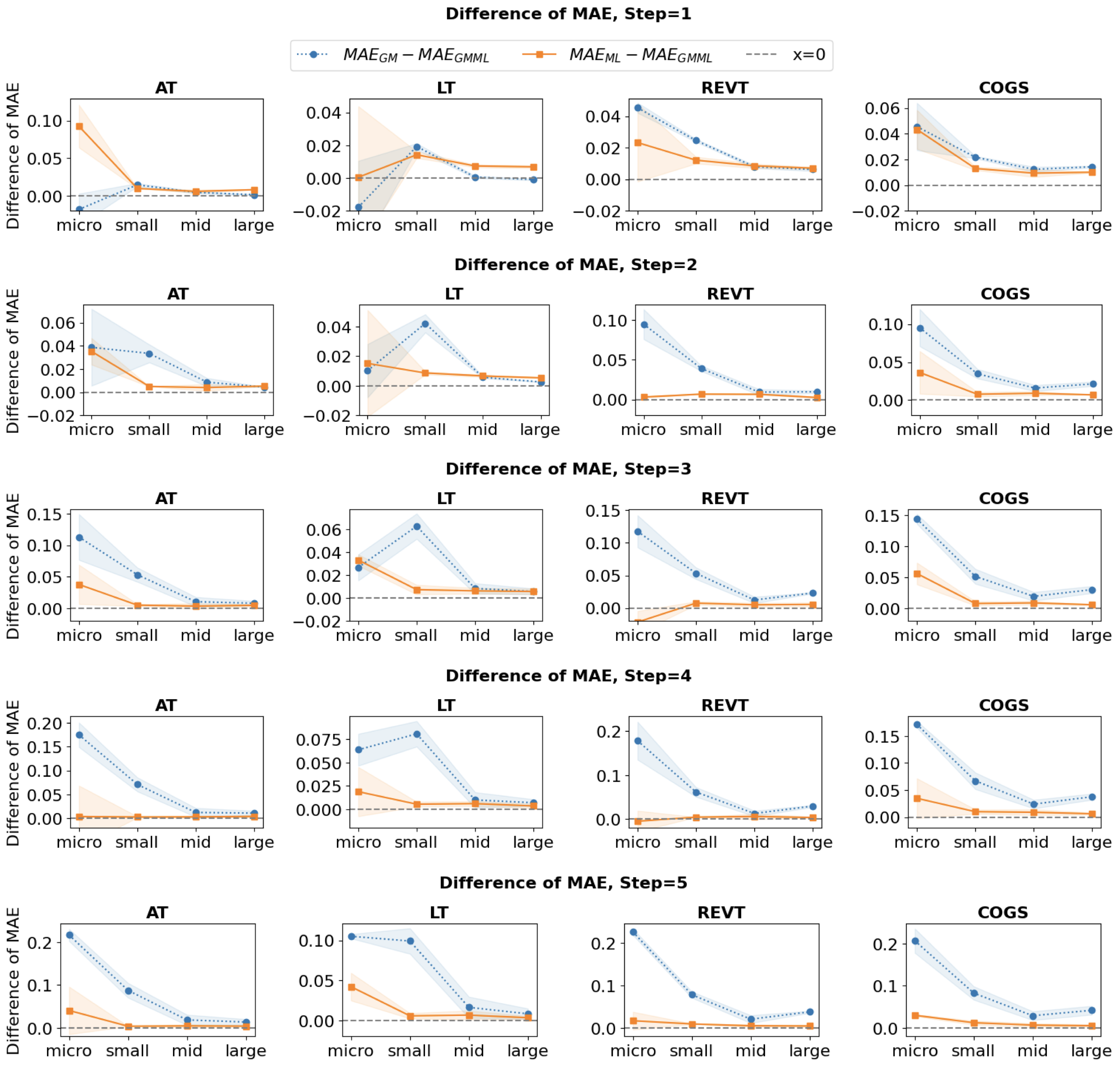}
    \caption{Visualization of performance gain across size groups for 5-step-ahead predictions}
 
    \label{fig:size_si}
\end{figure}

\begin{figure}
    \centering
    \includegraphics[width=1\linewidth]{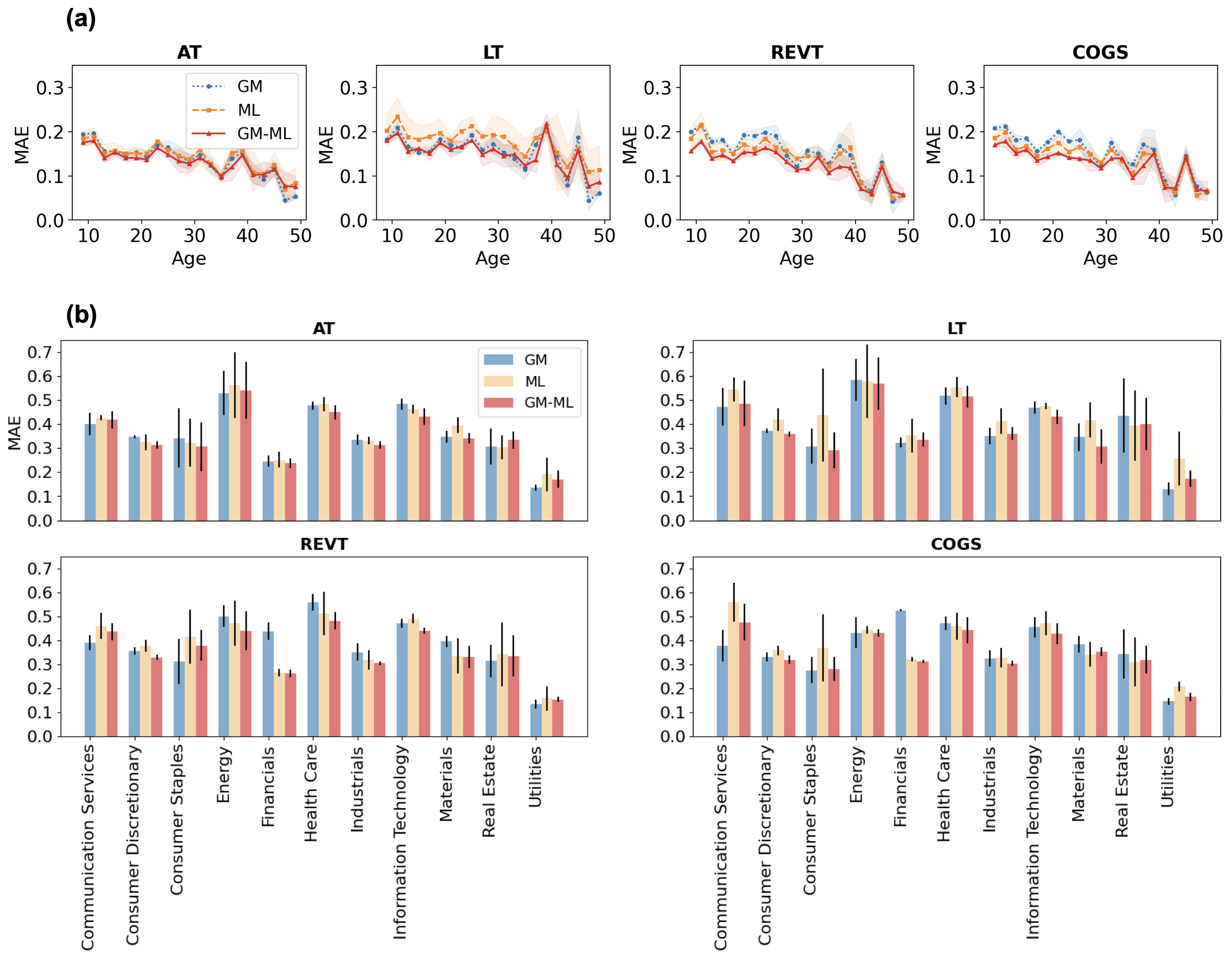}
    \caption{(a).Comparison of MAE across different age groups of companies for different models. GM is solid blue,  ML is dashed yellow and GM-ML is solid red line. The columns represent different prediction time steps. (b). Comparison of MAE across different sectors for GM-ML.\iffalse all these MAE errors are multi-step predictions that compute step=3.\fi}
    \label{fig:age}
\end{figure}


\end{document}